\documentclass{article}
\usepackage{spconf,amsmath,graphicx,subfigure, amsfonts}
\usepackage{enumerate}
\usepackage[english]{babel}
\usepackage{indentfirst}
\usepackage{adjustbox}

\usepackage{array}
\newcolumntype{\$}{>{\global\let\currentrowstyle\relax}}
\newcolumntype{^}{>{\currentrowstyle}}

\title{Adaptive DCTNet for Audio Signal Classification}
\name{Yin Xian*, Yunchen Pu*, Zhe Gan*, Liang Lu$^{\dagger}$ and Andrew Thompson$^{\mathsection}$}
\address{*Department of Electrical and Computer Engineering, Duke University, Durham NC 27708, USA \\
$^{\dagger}$Toyota Technological Institute at Chicago, Chicago IL 60637, USA \\
$^{\mathsection}$Mathematical Institute, University of Oxford, Oxford OX2 6GG, UK
}
\begin{document}
%\ninept
%
\maketitle

\begin{abstract}
In this paper, we investigate DCTNet for audio signal classification.  Its output feature is related to Cohen's class of time-frequency distributions. We introduce the use of adaptive DCTNet (A-DCTNet) for audio signals feature extraction. The A-DCTNet applies the idea of constant-Q transform, with its center frequencies of filterbanks geometrically spaced. The A-DCTNet is adaptive to different acoustic scales, and it can better capture low frequency acoustic information that is sensitive to human audio perception than features such as Mel-frequency spectral coefficients (MFSC). We use features extracted by the A-DCTNet as input for classifiers. Experimental results show that the A-DCTNet and Recurrent Neural Networks (RNN) achieve state-of-the-art performance in bird song classification rate, and improve artist identification accuracy in music data. They demonstrate A-DCTNet's applicability to signal processing problems.
\end{abstract}
\section{Introduction}
\label{sec:intro}
Learning feature representation of audio signals is one of the key interests for audio classification. Audio signals are rich in physical characteristics, such as energy, fundamental frequency, and formant, as well as in perceptual characteristics, such as pitch, timbre and rhythm~\cite{gerhard2003audio}, while they are usually contaminated by various kinds of noise. A good feature representation should represent those characteristics in compact forms, while robust to various kinds of noise.
Traditional audio signal features, such as Mel-Frequency Cepstral Coefficients (MFCC) and ERB-rate scale features can reveal the intrinsic attributes of audio signals. They are proposed based on auditory and physiological evidence of how humans perceive audio signals~\cite{stevens1937scale}, and are closely related to the short time Fourier transform or the constant-Q transform~\cite{davis1980comparison}. They are almost invariant under local scaling and frequency shift. However, the major drawback of these feature is that they are sensitive to noise~\cite{openshaw1994limitations}.

%reveal the
Recently, Convolutional Neural Networks (CNNs) have demonstrated great success in audio signal classification~\cite{lecun1995convolutional, hinton2012deep, lee2009unsupervised}. CNNs are able to carry out accurate parameter estimation~\cite{jain2009natural} even in the presence of significant noise. They employ learned filters, typically obtained through stochastic gradient descent (SGD) methods, and convolve them with signals to obtain features for classification~\cite{lecun1995convolutional}. However, they depend heavily on expert parameter tuning, and it is computationally expensive to learn the coefficients of the filters. It is also hard to physically interpret the learned coefficients, although we note that some studies found similarities of mel-filter banks and learned filter banks~\cite{sainath2013learning}.

On the other hand, the scattering transform, proposed by Mallat~\cite{mallat2012group}, employs a pre-specified collection of wavelets as filters, have obtained state-of-the-art results on some music and speech datasets~\cite{anden2014deep}. The structure of the scattering transform is similar to a cascade of constant-Q or mel-filter banks, and the scattering transform can capture useful spectral content of acoustic signals. The scattering coefficients can illustrate chord and attack interferences~\cite{anden2014deep}. A similar approach was proposed in the form of  PCANet~\cite{chan2014pcanet}, which uses eigenfunctions from eigen-decomposition as filters. PCANet has achieved competing results in image classification. Compared with CNNs, the filters of these two models are more interpretable, easier to learn, or learned for free.

Inspired by the scattering transform and PCANet, DCTNet was proposed in~\cite{xian2015whale, xian2016dctnet, ng2015dctnet}. DCTNet uses a prefix cosine function as filter; this is an approximation of the eigenfunction used in PCANet. DCTNet performs multilayer short time Discrete Cosine Transform (DCT). We investigate specially a two-layer DCTNet: a first DCT of the signal, and then a second DCT on each frequency level output of the first transform.
%At this stage we thus have many layer, with each layer a time-frequency representation in its own right.
Typical spectral feature representation for acoustic feature such as short time Fourier transform, spectrogram and linear frequency spectral coefficients (LFSC) can be related to each layer's output. As seen in~\cite{xian2016dctnet}, DCTNet output can improve classification accuracy on whale vocalizations.

In this paper, we further investigate the theoretical properties of DCTNet. One useful feature that can be computed from DCTNet is the sum over layers of the absolute value square of the DCTNet output. We show that this feature function is of Cohen's class of time-frequency distributions.

We also introduce two new extensions of the DCTNet framework which lead to state-of-the-art results for audio classification. Firstly, inspired by constant-Q transforms, we use geometrically spaced center frequencies, in order to make DCTNet adaptive to different acoustic perception scales.
We developed A-DCTNet for audio signals classification, it can better capture low frequency acoustic signal components that are sensitive to human than DCTNet and MFCC. %More technical details can be found from Sections~\ref{sec:dctnet} to Section~\ref{sec:mdctnet}.
Experiments on Artist20~\cite{ellis2007classifying} music data and bird song data show that, with a standard linear SVM classifier, features obtained from A-DCTNet improve classification rate over other features, including Mel-Frequency Spectral Coefficients (MFSC) and features obtained from the scattering transform.

Secondly, in order to make use of the sequential information of audio signals, and extract the higher level audio feature, we apply the Recurrent Neural Network (RNN) for further audio feature exaction and classification. State of the art classification accuracy is achieved in bird song data. This shows that the modified DCTNet is a simple flexible and effective feature extractor for audio signal processing.

\section{DCTNet}
\label{sec:dctnet}
DCTNet is essentially a multilayer short time DCT~\cite{xian2015whale,xian2016dctnet}. Similar to the short time Fourier transform, the short time DCT can be expressed as~\cite{oppenheim1978applications}:
\begin{align*}
X(m,k)= [x*(h\cdot c_k)](m)
\end{align*}
where $x$ is an input audio signal, $h$ is a window function, and $c_k$ are cosine functions. When $c_k$ are DCT-II functions,
\begin{align*}
X(m,k)=\sum\limits_{n=0}^{N-1}x(n)h(n+m)\cos\left(\frac{\pi}{N}(n+\frac{1}{2})k\right) ,
\end{align*}
where $N$ is the length of window, and $k=0,\cdots, N-1$. The process of short time DCT can be viewed as a modulation of bandpass filters to a signal.

\begin{figure}[htb]
\centering
\includegraphics[height=50mm,width=85mm]{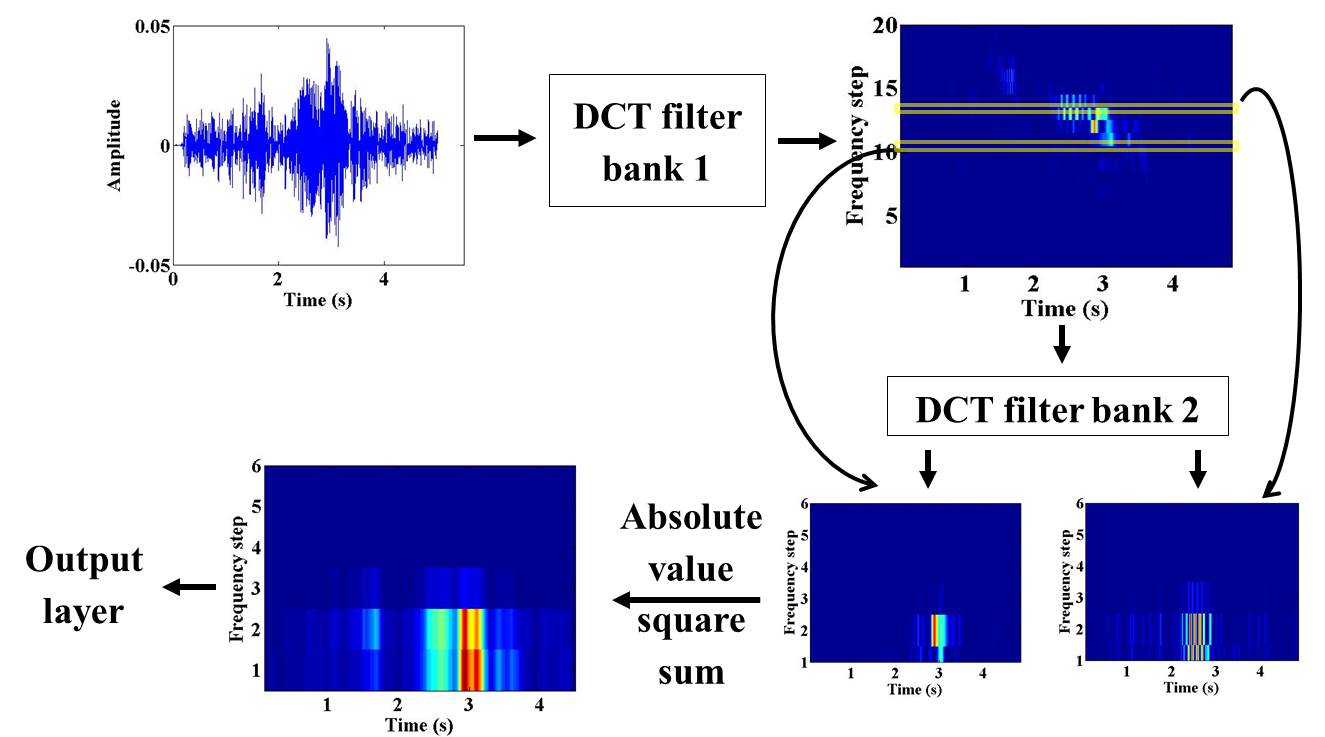}% Here is how to import EPS art
\vspace{-5pt}
\caption{%\small{
Two-layer DCTNet Process. The input is an oscillogram of an acoustic signal. After convolving the signal with DCT filterbanks, we have a short time DCT of the signal, which is a two-dimensional representation. Each row of the short time DCT is also a time series, convolving it with another DCT filterbanks, then we have a short time DCT inside a short time DCT. Summing the absolute value square of the second layer's outputs, we have a LFSC like feature of the signal, and we use it for classification.
%}
}
\label{fig:dctnet_process}
\end{figure}

A multilayer short time DCT can be expressed as:
\begin{align*}
X(m,k_1,\cdots,k_r)=[x*(h_1\cdot c_{k_1})*\cdots*(h_r\cdot c_{k_r})](m),
\end{align*}
where $r$ is the number of layers. Note that the whole operation is still linear. Choosing different window length in each layer can highlight detailed energy distribution of signals in the time-frequency plane, as shown in Figure~\ref{fig:dctnet_process}. Because the center frequencies in filterbanks are linearly spaced, the output of a two-layer short time DCT is closely related to the Linear Frequency Spectral Coefficients (LFSC) in acoustics. DCTNet has been shown in~\cite{xian2016dctnet} to improve classification accuracy of underwater acoustic signals. A simple illustration of a two-layer DCTNet is shown in Figure~\ref{fig:dctnet_process}.

A DCTNet variant is used for image feature extraction, applying 2D convolution, histograming and binary hashing~\cite{ng2015dctnet}. In contrast, we adopt DCTNet for acoustic signals, using an entirely different post-processing strategy~\cite{xian2015whale, xian2016dctnet}.

\section{Time-frequency analysis in DCTNet }
We first show that the output of the two-layer DCTNet is a time-frequency distribution of Cohen's class. From Section~\ref{sec:dctnet}, the two-layer DCTNet output is:
\begin{align*}
X(m,k_1,k_2)=[x*(h_1\cdot c_{k_1})*(h_2\cdot c_{k_2})](m).
\end{align*}

Summing the absolute value square of the second layer's output, the DCTNet feature function $F$ is:
\begin{align*}
F(m,k_2)=\sum_{k_1}|[x*(h_1\cdot c_{k_1})*(h_2\cdot c_{k_2})](m)|^2
\end{align*}
If we want to express a similar expansion in the continuous domain, for $\omega_1,\omega_2\in \mathbb{R}^d$, we pick $h_1,h_2\in \mathcal{S}(\mathbb{R}^d)$, $x\in\mathcal{S'}(\mathbb{R}^d)$. $\mathcal{S}(\mathbb{R}^d)$ is the Schwartz class, and elements in the dual space $\mathcal{S'}(\mathbb{R}^d)$ of $\mathcal{S}(\mathbb{R}^d)$ are called tempered distribution~\cite{grochenig2013foundations}.  $x, h_1, h_2$ are real, and $h_1, h_2$  are even functions. The process can be expressed as
\small
\begin{align*}
&F(t,\omega_2)
=\int_{\mathbb{R}^d} \left|x(t)*(h_1\cdot c_{\omega_1}(t))*(h_2\cdot c_{\omega_2}(t))\right|^2 d\omega_1
\end{align*}
\normalsize
where $c_{\omega}(t)=\cos(2\pi\omega t)$. $c_{\omega_1}$ and $c_{\omega_2}$ are cosine functions of the first layer and second layer respectively, and $h_1$ and $h_2$ are window functions of the first layer and second layer respectively.  Expanding the above expression, the cosine function integrates to a delta function. Since a cosine function can be expressed as a summation of two exponential terms, the above expression can be simplified as
\small
\begin{align}
F(t,\omega_2)=&\int\int_{\mathbb{R}^{2d}} x(t-u)x(t-u')\phi(u,u') e^{-j2\pi\omega_2(u-u')}dudu' \notag \\
 &+ \text{its complex conjugate},
\label{eq:dctnet_2layer}
\end{align}
\normalsize
where  $\phi(u,u')=C\cdot\int_{\mathbb{R}^{d}}(h_1(\tau))^2 h_2(u-\tau)h_2(u'+\tau)d\tau$ is a kernel function, and $C$ is some constant.

Applying a change of variables to the right hand side in the above formula: $t_1=t-u'$ and $t_2=t-u$, we have
\small
\begin{align}
&\int\int_{\mathbb{R}^{2d}} x(t_1)x(t_2)\phi(t-t_1,t-t_2)
e^{-j2\pi\omega_2(t_1-t_2)}dt_1dt_2 \notag \\
=&\int\int_{\mathbb{R}^{2d}} x(t'+\frac{\tau}{2})x(t'-\frac{\tau}{2})\psi(t-t',\tau)e^{-j2\pi\omega_2\tau}dt'd\tau ,
\label{eq:dctnet_2nd}
\end{align}
\normalsize
where the kernel function is
$\psi(t_1,t_2)=\phi(\frac{-t_1-t_2}{2},t_1-t_2)$~\cite{o1999shift}.
Eq.~(\ref{eq:dctnet_2nd}) is of Cohen's class of time-frequency distributions~\cite{cohen1995time}. The Cohen's class of time-frequency distributions are shift covariant. This means that they are especially amenable to building invariant features by means of pooling.

Note that, given the linearity of scale of DCT filterbank, the output of the second layer of DCTNet gives a linear frequency spectrogram like feature~\cite{xian2016dctnet}.

\section{Adaptive DCTNet (A-DCTNet)}
\label{sec:mdctnet}
We extend DCTNet by making it adaptive to different acoustic perception scales. In the constant-Q transform~\cite{brown1991calculation}, the center  frequencies are logarithmically spaced. Such spacing is similar to the human auditory perception system~\cite{seneff1985pitch, daubechies1990wavelet}. The constant-Q transform has shown to be useful in audio signal processing~\cite{schorkhuber2010constant}.

%Constant-Q transform is originated from discrete Fourier transform (DFT), and is used initially in musical signal processing. The constant-Q transform
We define the adaptive short time DCT as
\begin{align}
%X_{CQT}(m,k)&=\sum\limits_{n=0}^{N[k]-1}h_k(n+m)x(n)\cos\left(\frac{\pi Q}{N[k]}(n+\frac{1}{2})\right) \\
X_{CQT}(m,k)=\sum\limits_{n=0}^{N-1}x(n)h_k(n+m)\cos(\pi f_k(n+\frac{1}{2})/f_s)
\end{align}
where $f_s$ is the sampling frequency, $f_k$ is a geometrically spaced centered frequency given by
\begin{align}
f_k=f_0\cdot 2^{\frac{k}{b}},\qquad k=1,2,\cdots,K,
\end{align}
$b$ is the number of filters per octave, and K is the overall number of frequency bins. In the constant-Q transform, the Q factor, namely the ratio of center frequency to bandwidth, is constant.
The modified DCTNet is a multilayer modified short time DCT.

%The constant-Q transform closely related to the wavelet transform~\cite{daubechies1990wavelet}. In scattering network, it computes a cascade of wavelet convolution and moduli~\cite{}

\begin{figure}[!ht]
% \hfill
\centering
\subfigure[Log-MFSC feature]{\includegraphics[width=.49\linewidth,height=3.3cm]{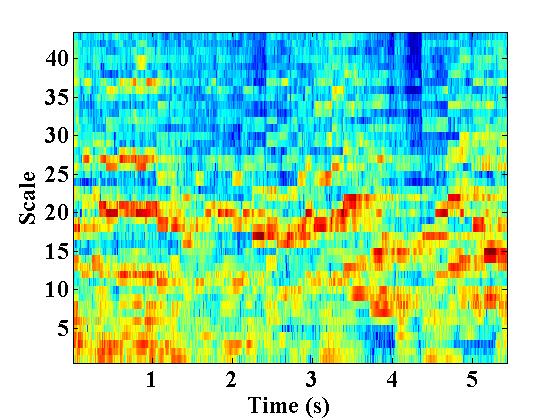}}
% \hfill
\subfigure[The $2nd$ layer output of the A-DCTNet]{\includegraphics[width=.49\linewidth,height=3.3cm]{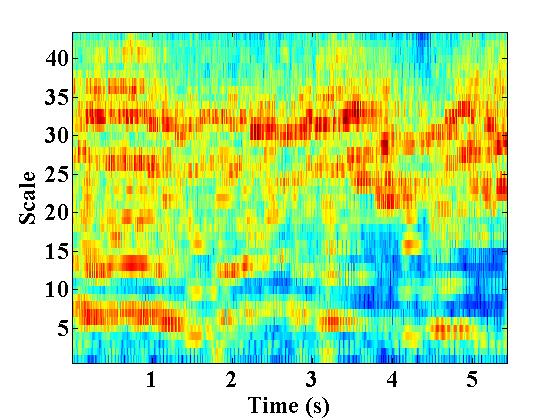}}
\caption{Comparison of Log-MFSC feature and A-DCTNet second layer output.}
\label{fig:mdctnet}
\end{figure}

With the frequency bins geometrically spaced, the two-layer adaptive DCTNet has the output shown in Fig~\ref{fig:mdctnet}.
We use as an example a piece of classical music by Bach. Choosing $b=12$ for the first layer, and $b=6$ for the second layer. By adjusting the window function, the low frequency signal components in A-DCTNet can be better captured than that of MFSC in the plots.

\section{Recurrent Neural Networks}
\label{sec:rnn}
\begin{figure}[htb]
\centering
\includegraphics[height=40mm,width=90mm]{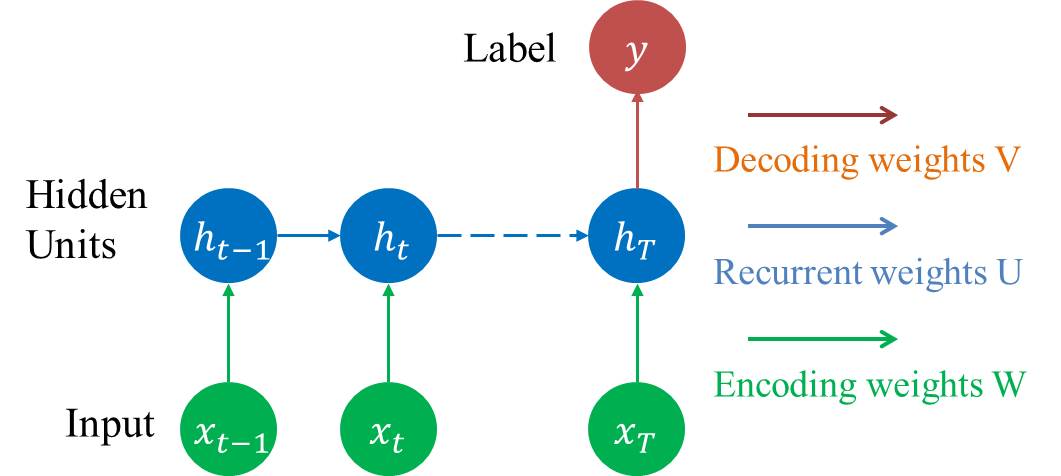}% Here is how to import EPS art
\vspace{-5pt}
\caption{Process of RNN. $\boldsymbol{x}$ is the input sequential data, $\boldsymbol{h}$ is the hidden units, and $\boldsymbol{y}$ is the label.}
\label{fig:rnn_process}
\end{figure}

We further extend A-DCTNet by combining it with  Recurrent Neural Networks (RNN)~\cite{hochreiter1997long,cho2014learning} in audio classification task. RNN are known to be  powerful models for capturing features of sequential data. The process of RNN is shown in Figure~\ref{fig:rnn_process}.
In the figure, $\boldsymbol{x}=(x_1,x_2,\cdots,x_T)$ is an input sequence, $x_t$ is an input at time step $t$. $\boldsymbol{h}=(h_1,h_2,\cdots,h_T)$ is a hidden unit, $h_t$ is a hidden state at time step $t$, and $\boldsymbol{y}=y_T$ is the output. We recursively update the hidden unit state by $h_t=g(h_{t-1},x_t)$, where $g(\cdot)$ is an activation function. We predict $\boldsymbol{y}$ from $\boldsymbol{x}$ through $\boldsymbol{h}$: $p(\boldsymbol(y|x)=p(y_T|h_T))$. Feature labels are then determined by $p(y)=\text{softmax}(V\cdot h_T)$.

\vspace{-3pt}
\section{Experimental results}
\label{sec:results}

\subsection{Datasets}
\label{ssec:data}
We use the Artist20 music data~\cite{ellis2007classifying} and bird song data collected by the Nowicki Lab at Duke University for experiments. Artist20 is a database of six albums by 20 artists. Different artists have their own styles of performing and singing.
Artist20 has 1413 tracks, and each track is 30 seconds in length. The sampling frequency of the original data is 44kHz. Our goal is to identify the artists from the music.

For the bird song data, there are four types of bird songs, and there are 200 tracks in total. The sampling frequency of the data is 44kHz. Each piece of bird song lasts about 2-3 seconds. Our goal this time is to classify this bird songs.

\vspace{-4pt}
\subsection{Classification results}
\label{ssec:results}
\vspace{-4pt}
We use features obtained from the two-layer A-DCTNet and the scattering transform, and we make a comparison with MFSC, LFSC and ERB-rate scale features, applying the linear SVM and RNN to classify the data. In the two-layer A-DCTNet, we let the first layer and the second layer maximum frequency be 5500Hz, and minimum frequency be  40Hz. The major signal energy components are distributed in the 40Hz to 5500Hz region. For the scattering transform feature, we use the MATLAB toolbox~\cite{scattering_toolbox} with time frame duration $T=125ms$ (or window size 256), and set $Q_1=8$, $Q_2=1$ to best capture the acoustic signal information~\cite{anden2014deep}. For MFSC, LFSC and ERB rate scale features, we use a Hamming window of length 256, and step size 128 to create the spectrogram, then multiply the spectrogram with the filter bank function~\cite{voicebox}. We extract 40 coefficients from each time frame, and concatenate the coefficients along the time axis.

We use 80\% of the music data and bird song data for training, and the rest for testing. The accuracy of different features sets for classification of the Artist20 and bird song data are shown in Table~\ref{table:Accuracy1} and Table~\ref{table:Accuracy4}, respectively. The baseline accuracy for the Artist20 identification is 54.49\%~\cite{ellis2007classifying}, which uses MFCC as features, and applies a Hidden Markov Model (HMM) and a Gaussian Mixture Model (GMM) for classification.

\begin{table}[t!]
%\small
\caption{Accuracy (\%) of different feature sets for music} %\label{table1}%title of the table
\label{table:Accuracy1}
%\vspace{1pt}
\centering % centering table
%\begin{adjustbox}{minipage=1.05\linewidth,scale=0.98}
	\begin{tabular}{ccc}
	\hline%\hline
	Feature set & linear SVM (\%) & RNN (\%) \\
	\hline
	A-DCTNet $2^{nd}$ layer & $\bold{45.23}${\scriptsize$\pm 1.34$}  & $\bold{79.11}${\scriptsize$\pm 0.11$}  \\
	Scattering transform &  $40.59${\scriptsize$\pm 1.24$} & $71.24${\scriptsize$\pm 0.21$} \\
	MFSC & $32.16${\scriptsize$\pm 1.43$} & $75.09 ${\scriptsize$\pm 0.21$} \\
	LFSC & $33.17${\scriptsize$\pm 1.27$} & $74.12${\scriptsize$\pm 0.24$} \\
	ERB scale & $16.34${\scriptsize$\pm 2.52$} & $51.21${\scriptsize$\pm 0.47$}
	\\ %[0.5ex]
	\hline %inserts single line
	Baseline: MFCC+HMM & 54.49 &
	\\ %[0.5ex]
	\hline %inserts single line
	\end{tabular}
%\end{adjustbox}
% is used to refer this table in the text
\end{table}

With linear SVM, A-DCTNet features improve classification rate over other features. However, without using the sequential information of audio signals, the overall classification performance is not desirable for music data. In order to improve classification accuracy, we incorporate the use of RNN.

%\vspace{-5pt}

We use RNN with one hidden layer, and initialize all recurrent matrices with orthogonal initialization. Non-recurrent weights are initialized from a uniform distribution in $[-0.01, 0.01]$. The experiments are implemented with Theano~\cite{Bastien-Theano-2012}. When applying RNN, we divide each track of music and bird song into small chunks. For music data, the length of each chunk is 60, with overlap size 30. For the bird song data, the length is 40 with overlap size 20.

The A-DCTNet achieves state-of-the-art results for bird song classification, and improves identification accuracy for the music Artist20 dataset. The two-layer A-DCTNet outperforms all other methods using both classifiers, due to its increase ability to capture low-frequency acoustic information. The main reason for the improved performance with the RNN classifier is the capturing of sequential information inherent in the signals. One possible further explanation for the improve performance over the scattering transform is that phase information is preserved in each layer of A-DCTNet. In contrast, in the scattering transform, absolute values are taken after each convolution in order to achieve translation invariance.

\begin{table}[t!]
	%\small
	\caption{Accuracy (\%) of different feature sets for bird songs} %\label{table1}%title of the table
	\label{table:Accuracy4}
	%\vspace{1pt}
	\centering % centering table
	 %\begin{adjustbox}{minipage=1.05\linewidth,scale=0.98}
	\begin{tabular}{ccc}
		%\begin{tabular}{|$l|^l|^l|}
		\hline%\hline
		Feature set & linear SVM (\%) & RNN (\%) \\
		\hline
		%\rowstyle{\bfseries}
		A-DCTNet $2^{nd}$ layer & $\bold{95.26}${\scriptsize$\pm 1.34$}  & $\bold{98.65}${\scriptsize$\pm 2.21$}  \\
		Scattering transform &  $91.42${\scriptsize$\pm 2.24$}  & $94.64${\scriptsize$\pm 2.42$}  \\
		MFSC & $92.91${\scriptsize$\pm 1.43$}  & $95.26${\scriptsize$\pm 2.43$}  \\
		LFSC & $90.17${\scriptsize$\pm 2.77$}  & $95.12${\scriptsize$\pm 2.22$}  \\
		ERB scale & $86.34${\scriptsize$\pm 2.22$}  & $92.73${\scriptsize$\pm 2.31$}
		\\ %[0.5ex]
		\hline %inserts single line
	\end{tabular}
	%\end{adjustbox}
	% is used to refer this table in the text
\end{table}
\vspace{-6pt}

\section{Conclusion}
\label{sec:conclude}
\vspace{-4pt}
In this paper, we analyze the time-frequency characteristics of DCTNet, and show that the output of a two-layer DCTNet is of Cohen's class of time-frequency distributions. We propose an A-DCTNet which uses the idea of the constant-Q transform, to make it adaptive for different acoustic perception scales. With a standard linear SVM classifier, the A-DCTNet improves classification accuracy over other features, including MFSC and the scattering transform. The classification accuracy can be further improved by using RNN, which takes audio sequential information into consideration. With the use of RNN, A-DCTNet achieves state-of-the-art in bird song classification, and significantly improves classification accuracy in music artist identification. Our results demonstrate that A-DCTNet is a simple, flexible and effective feature extractor for audio signal processing.

\vspace{-6pt}
\section{Acknowledgement}
\label{sec:acknowledge}
\vspace{-4pt}
We thank Dr.~Ingrid Daubechies for great instructions and discussions. We thank Dr.~Steve Nowicki for providing dataset, and Dr.~Xiaobai Sun, and professors and friends of "Mathematics of Signal Processing" program at Hausdorff Institute of Mathematics for suggestions and comments. We thank Dr.~Loren Nolte, Dr.~Larry Carin and Dr.~Robert Calderbank for support and help.

% Below is an example of how to insert images. Delete the ``\vspace'' line,
% uncomment the preceding line ``\centerline...'' and replace ``imageX.ps''
% with a suitable PostScript file name.
% -------------------------------------------------------------------------

%\vfill
%\pagebreak

% \vfill\pagebreak

\bibliographystyle{IEEEbib}
\bibliography{refs}

\end{document}